# MT4j – A Cross-platform Multi-touch Development Framework

**Uwe Laufs**
Fraunhofer IAO
Nobelstraße 12
70569 Stuttgart
uwe.laufs@iao.fhg.de

**Christopher Ruff**
Fraunhofer IAO
Nobelstraße 12
70569 Stuttgart
christopher.ruff@iao.fhg.de

**Jan Zibuschka**
Fraunhofer IAO
Nobelstraße 12
70569 Stuttgart
jan.zibuschka@iao.fhg.de

**ABSTRACT**
This article describes requirements and challenges of cross-platform multi-touch software engineering, and presents the open source framework Multi-Touch for Java (MT4j) as a solution. MT4j is designed for rapid development of graphically rich applications on a variety of contemporary hardware, from common PCs and notebooks to large-scale ambient displays, as well as different operating systems. The framework has a special focus on making multi-touch software development easier and more efficient. Architecture and abstractions used by MT4j are described, and implementations of several common use cases are presented.

**Categories and Subject Descriptors**
H.5.2 [**User Interfaces**]: Input Devices and Strategies; D.2.11 [**Software Architectures**]: Domain-specific Architectures; D.2.7 [**Distribution, Maintenance, and Enhancement**]: Portability;

**INTRODUCTION**
Multi-touch technology has received considerable attention in recent years. Even though the technology has existed for more than 20 years, the current hype regarding the multi-touch interaction started no earlier than 2005, when an article about low-cost construction of multi-touch screens [14] made multi-touch technology available to a bigger community. In 2006, a video [8] showing multi-touch interaction experiments and potentials of the technology was hyped across the internet, which was the first contact with the technology for many consumers.

In the meantime, products with multi-touch capabilities are readily available and very successful, for instance Apple's iPhone [1] or Microsoft's interactive table Microsoft Surface [19]. It is also likely that Microsoft's support for multi-touch in Microsoft Windows 7 [16] will further increase the relevance of the technology [25].

Even though there is currently a lot of activity in research, industry and in the open source community, there are still barriers on the technical side regarding the development of multi-touch applications in practice, as there are few frameworks supporting the cross-platform development of multi-touch applications. Resulting from this gap, there is a contemporary development towards the provision of more comprehensive frameworks supporting multi-touch interaction, and lowering implementation barriers across a variety of hardware.

Multi-touch for Java (MT4j) is an open source multi-touch application development framework which is released under the GPL license and can be freely used by anyone. The framework provides high level functionality and aims at providing a toolkit for easier and faster development of multi-touch applications.

**RELATED WORK**
Multi-touch has a long and eventful history, starting as early as the beginning of the 1980s. There have been various instantiations of the technology, differing in form factor (ranging from small to huge displays), device integration (ranging from PDA to ambient screens), and application fields (such as CAD, robotics, and entertainment), as well as targeting user groups from children [21] to the elderly [9]. A comprehensive survey of multi-touch history is given by [11].

Mirroring this history, there are a multitude of programming APIs related to multi-touch systems. However, those solutions seldom reach the status of portable, reusable frameworks. Such solutions are mostly realized as either application-specific architectures or reusable components for specific low-level problems, and are provided without further abstraction/integration effort. This problem is exuberated by the variety of non-standardized multi-touch hardware [15]. During the last years, there has been a development towards the provision of more comprehensive frameworks supporting multi-touch interaction, lowering implementation barriers.

The Squidy [17] library offers a high level abstraction for device- and platform-independent input processing that is quite similar to the implementation the lower layers of MT4j, but does not offer platform-independent UI components.

pyMT [2], is based on the Python programming language. It is portable and offers a broad framework similar to the one presented here. However, compared to our solution, the performance requirement is not met as reliably (which is a common issue of many python-based solutions).

Microsoft Windows 7/WPF offer a comprehensive API for multi-touch support based on the WM_TOUCH message. While this framework has a very nice performance, cross-platform portability is of course an issue. The integration of multi-touch technology in Microsoft Silverlight 3 [20] has the potential to remedy this problem, but is not applicable to open source multi-touch solutions (yet), as Silverlight 3 is not supported there.

Similarly, Adobe Flash/Air also has begun to integrate multi-touch support, but here, there are severe performance issues, as full screen updates are notoriously performance intensive in Flash. Work to fix this is also ongoing, and the latest Flash Player 10 brings some improvements, at least in the area of full screen videos [3].

## REQUIREMENTS FOR MULTI-TOUCH FRAMEWORKS

In order to reduce implementation efforts during multi-touch software development, there are several requirements and challenges which can be addressed by a development framework in a generic and reusable way.

### Portability

In the market, there are a lot of operating systems and multi-touch input protocols such as WM_TOUCH (the native interface for multi-touch hardware in Microsoft Windows 7) or the TUIO-protocol [15] which has gained the status of a de facto standard protocol especially among open source multi-touch solutions.

To be truly portable, a multi-touch framework needs to offer an abstraction supporting all those environments.

### Input Abstraction

Most multi-touch applications intensively use multi-point gestures like zooming the screen content or rotating objects with two fingers. Multi-point gestures are gestures that include more than one input motion. The realization of such gestures also requires the capability to process multiple input motions on the software side. In addition, multi-touch input systems are often integrated with additional input options, such as digital pens [18][22].

Because different applications may require additional gestures or use gestures differently than expected, gesture recognition and gesture processing has to be flexible and extensible.

### Performance

Multi-touch applications often have a special focus on user experience and graphically rich user interfaces. Also, unlike mostly static user interfaces, the usage of gestures like the zoom gesture on full screen content requires a complete redraw of the whole screen. In such cases, common optimization strategies, like updating only changed parts of the user interface can often not be applied.

To achieve the required frame rates for smooth and direct user interface interaction, there is a necessity for high graphics rendering performance in many cases.

### UI Integration

Traditional user interfaces commonly use input devices like a mouse and a keyboard. Many of these applications are based on a windowing system and make use of a windowing toolkit. Existing windowing toolkits already provide mechanisms for handling user interaction and provide a wide range of UI components and widgets which can be composed to complex user interfaces. Unlike a mouse, a multi-touch screen can provide more than one input motion at the same time. A mouse can provide additional events (wheel, buttons) instead, which cannot be directly produced using a multi-touch screen. Having multiple input motions causes problems regarding the use of windowing toolkits because these toolkits normally support a single pointer device only. Additionally, to reap the full benefits of a multi-touch implementation, new UI components based on multi-touch interaction concepts are needed.

To reduce implementation times and offer appropriate interaction models, a toolkit of multi-touch UI components should be provided.

## THE MT4j APPROACH

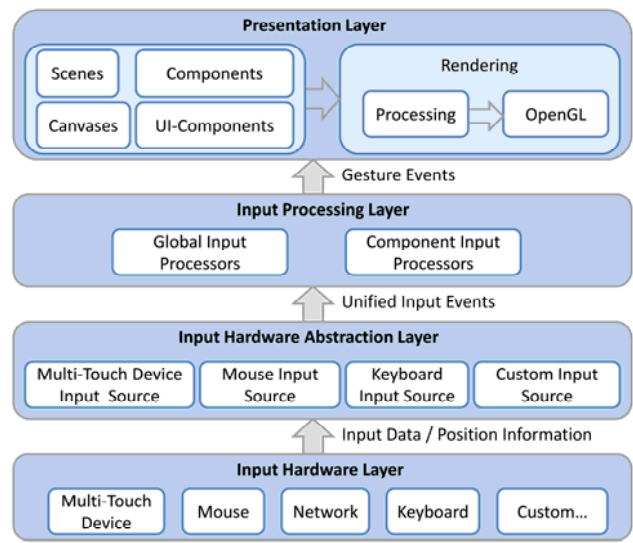

Figure 1. MT4j architecture overview

The functionality of the MT4j framework architecture is subdivided into different layers communicating through events sent from one layer to the next, as illustrated by Figure 1. The emphasis on input layers represents the importance of a flexible input architecture, while performance issues are mainly addressed by the presentation layer. This can be seen as a model-independent reduced MVC paradigm, with the input layer representing the controller, the presentation layer representing the view. The different layers are described in the next sections.

### Input Hardware Abstraction

By using a hardware abstraction layer, MT4j can support various input hardware with only minimal adjustments in the input hardware abstraction layer. In this abstraction layer, the different raw input data is converted into unified input events. The only step to be taken in order to support a new type of input hardware is to extend the abstract super class of all the input sources and add the functionality specific to this type of input. By providing this easy input extensibility, MT4j lends itself to many different applications and especially to areas where new kinds of human computer interaction methods and techniques are explored and applied.

MT4j comes with a set of implemented input providers including mouse, keyboard and multi-touch input protocols such as WM_TOUCH and the TUIO-protocol. Additionally, MT4j supports the use of multiple mice input on Microsoft Windows and Linux platforms, which facilitates testing of multi-touch functionality even without multi-touch capable hardware available. All of these input sources can be used synchronously and in combination without the risk of non-deterministic behavior.

**Input Processing**

The aspects of processing, analyzing and interpreting user input are very important for a framework focused on multi-modal input. In MT4j, the input processing occurs at two different stages in the input event flow. The first stage is the global input processing stage where a number of input processors can be registered which subsequently listen directly to the input of the various input sources. This stage is used when all input has to be processed. It also allows modification of user input before it is passed up to the next layers. For example, every newly created scene in MT4j (see 3.3) automatically registers a global input processor which checks if there is a component at the position of the input and sets that component as the target of that input event.

The second input processing stage is located at the component level. It allows processing of input that was targeted at one component only. Here, multi-touch gestures like the rotate and scale gestures can be found. These component input processors can be registered modularly with any component allowing for a pluggable behavior changeable at runtime. As the field of multi-touch interaction and gestures is currently an active field of research, it is understood that both global and component input processors have to be extensible and customizable. If the criteria for a multi-touch gesture are met, the input processor fires a gesture event carrying the information about the recognized gesture to the corresponding component which passes the event on to its gesture listeners.

The action taken when a gesture event is received is determined by the attached gesture listeners which can modify the component's behavior or appearance. For easy usage, the framework already includes default implementations of such listeners to react to the included gestures. The use of the observer pattern [13] in this way is a familiar concept to most Java programmers and is widely used in Java windowing toolkits.

Figure 2 shows the event flow from the input hardware abstraction layer to the input processing layer and finally to the presentation layer (from left to right). First, input events are produced by the underlying hardware. An input source listens to the events produced by the hardware (e.g. a multi-touch screen that sends motion events via the TUIO protocol). Based on these events, unified input events are produced and passed to the input processors on the input processing layer. These events are used to provide input specific functionality (e.g. a cursor is shown at every position where a multi-touch screen is currently touched) by global input processors. Component input processors translate the input events to higher level gesture events. For example, a rotate processor recognizes when a user rotates at least two fingers on top of an object and produces a rotate gesture event. Finally, the target components (user interface components on the presentation layer) listen to higher level gesture events and react by performing the desired action (e.g. a rotate event is used by the default rotate action to rotate the target object in the user interface).

**Presentation**

The presentation layer is a vital part of any application using a graphical user interface. Multi-touch applications are often created for specialized use cases that differ a lot from everyday office applications. Often, one of the requirements is the creation of an exciting and eye-catching user experience. This implicates, that the user interface elements provided by conventional GUI toolkits are often not appropriate. Instead, MT4j provides a flexible way to create customizable and media rich user interfaces. For this purpose, MT4j contains graphical components ranging from graphical primitives to more complex user interface widgets and is inherently designed to allow development of 2D and 3D applications.

In order to organize different aspects of an MT4j application, the concept of "scenes" was introduced. Scenes encapsulate and cleanly separate the input processing and presentation of one aspect of an application from another. An example of using scenes would be a game that has a menu scene and a game play scene which contain different business logic and interfaces. To navigate between scenes, a scene change can be triggered. The root component of every scene in MT4j is the canvas component. It acts as the link between the global input processing layer and the presentation layer. All input events pass through the canvas component, which then further propagates events to their destinations. It also contains methods for checking which components are located at a specified screen position and it is responsible for recursively drawing the canvas with all its child elements.

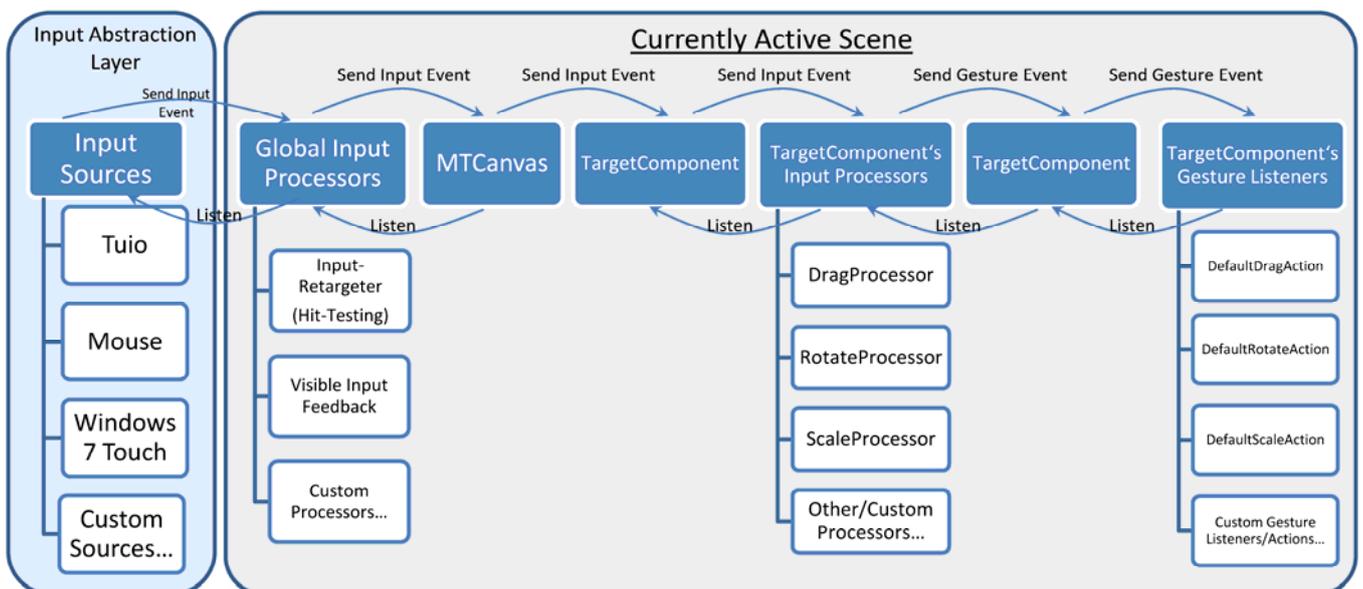

Figure 2. Input event propagation in MT4j

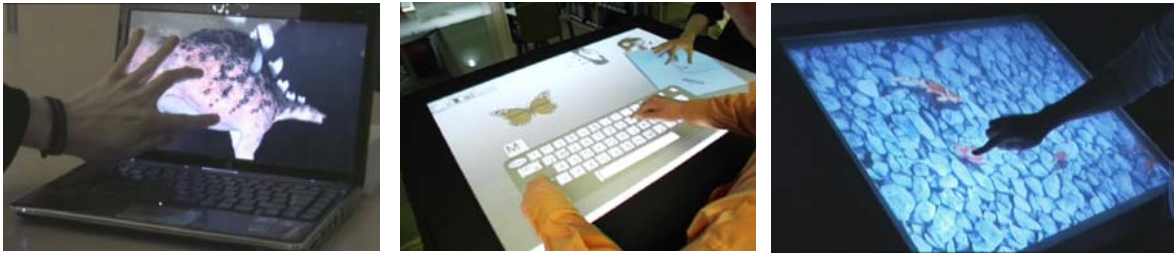

Figure 3. MT4j on different platforms (left to right: Windows 7, TUIO, proprietary/MIM).

In MT4j, graphical user interfaces are based on a hierarchic component structure which allows the composition of user interface components in a tree structure. Even complex composite components can interact with user gestures as a whole. This means that for example a rotate gesture can be applied to a component including all children within its component hierarchy.

There are two general types of components in MT4j. Invisible components provide basic functionality like the composition of components. Visible components can directly interact with the user input. Included visible components are primitive shape components (e.g. rectangle, polygon, ellipse and line) as well as a set of more complex user interface components. More complex user interface components are often based on primitive shapes and provide functionality like image rendering with support for common image file formats, rendering of scalable vector graphics [4] or rendering of 3D models. Also, components like buttons or vector based text rendering components are included in MT4j.

The component based structure of user interfaces in MT4j is extensible. When creating a custom user interface component, a lot of functionality can be reused from available components. Custom components can be built upon already available functionality (e.g. hit detection, gesture processing) by composing primitive shapes and available user interface components to more complex components. An example of such a higher level component is the multi-touch keyboard, which is composed of a wide range of MT4j user interface components.

For rendering of MT4j components, the processing [Processing 2009] toolkit is used. Processing is an open source Java toolkit aimed at the creation of data visualizations, interactions and computational art and has a very active community. It provides an easy syntax for accessing drawing and visualization functionality and contains many useful utilities. By using a rendering abstraction layer, it is possible to choose between different renderers. Software- and hardware accelerated renderers are available [5]. A disadvantage of the rendering abstraction is that not all available hardware accelerated functionality and features are exposed by the rendering API, which can prevent the use of optimization techniques like Vertex Buffer Objects or display lists. The underlying OpenGL context can also be accessed and used directly when necessary. When the OpenGL renderer is chosen, most MT4j components use this direct access for speeding up rendering, especially of complex components or 3D models by orders of magnitude. This allows for a very good performance of MT4j applications on newer systems, but also allows falling back to software rendering if run on older hardware.

## APPLICATION CASES

This section presents several application cases realized by ourselves and others using MT4j, illustrating versatility and viability of the framework, and also giving an impression of the increased implementation speed that can be realized using the MT4j framework. Additionally, it may serve as an overview of real-life multi-touch application cases that we came in contact with during the lifetime of the MT4j framework, illustrating that the technology is by no means lacking in that respect, as has been suggested by earlier research [23]. The cases presented here are just a short overview. MT4j has been used in far more application fields than can be illustrated here. We will present one application from a research project, one demo showcasing the capabilities of the framework, and third party applications demonstrating versatility and uptake of the framework.

### Collaborative Disaster Management

Based on MT4j, we are implementing a collaborative geo-information system aimed at planning and coordination of emergency response measures in the context of large public events in the context of the German public security project VeRSiert [6] (see Figure 4). While there is substantial earlier work in the area of multi-touch in emergency management [12] [24] [26], the event management domain has specific circumstances that make deployment of a multi-touch system in this case more viable than in classic emergencies, such as fires, floods or earthquakes.

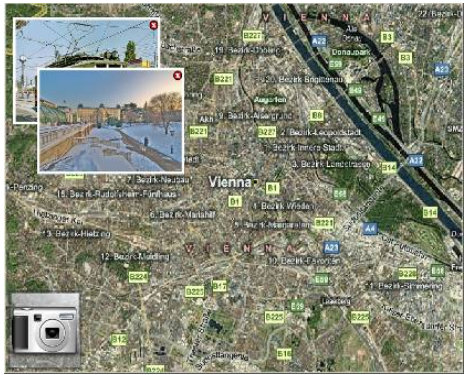

Figure 4. Collaborative Disaster Management.

The event scenario has stakeholders that are involved in planning and realization of many events, and are experts in the emergency management domain (e.g. police, fire brigade, and other public security forces). It also has a distinguished planning phase, during which the multi-touch implementation provided can be used to communicate information about street blocks, construction site, or available forces. Additional information aimed at non-expert stakeholders (such as operators or private security services) can be added to the map as annotation, along with operation manuals giving instructions for emergency cases.

**Flickr showcase**

The Flickr photo application (Figure 5) is a multi-user multi-touch application demonstration showcasing UI components, usability concepts, and web integration provided by the MT4j framework. It uses the MT4j keyboard component (representing the most desirable UI component in multi-touch systems according to a recent survey [10]) to allow the users to type in photo search terms. The keyboard is extended by a custom flickr button which triggers a photo search. When the button is pressed, the entered search term is passed over to the integrated Flickr interface [7] which returns the best matches to the application and starts to download and display the photos. For image sorting and manipulation, MT4j's default gesture processors and gesture actions are used to make photos movable, rotatable, scalable, groupable and to provide zoom and pan functionality for the full screen content.

**Third-party applications**

In addition to the development done by us in context of research and industry projects, third party developers are also actively putting MT4j to use to realize applications in several application fields, such as musical instrument emulators and UML editors (see Figure 6).

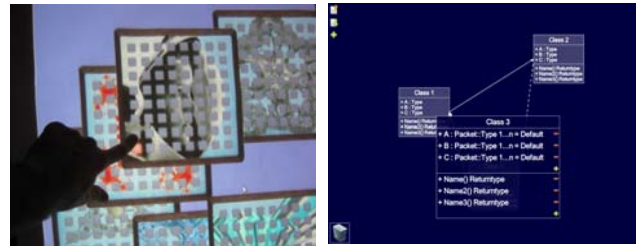

Figure 6. Third-party applications based on MT4j (Monome emulator, UML editor).

## DISCUSSION

The interest and positive feedback from the community, illustrated by more than 100.000 hits on MT4j.org per month (even more when new versions are released), and first third party implementations based on the framework, see previous section, show that there is a high demand for a multi-touch application development framework based on Java programming language. For the further development, it is intended to include contributions from the open source community to provide an even more comprehensive toolkit including - but not limited to - an extension of the UI module of the presentation layer providing a wide range of high level user interface components.

## CONCLUSION

In this paper we described critical requirements for cross-platform multi-touch software development frameworks and showed how the open source framework MT4j addresses these problems. The main focus of this article was architecture and design of the MT4j framework. In the showcases, we demonstrated examples of applications built with MT4j. During the development with MT4j, a lot of the provided functionality could be reused and implementation efforts were reduced. Currently, we use MT4j as platform for the development of multi-touch applications with a focus on multi-user business applications. This includes applications for collaborative planning and modeling. Additional components developed during these projects will be part of future MT4j releases.

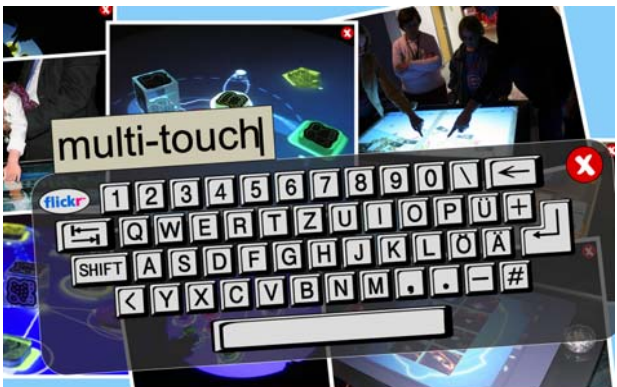

Figure 5. MT4j Flickr photo search showcase.

**BIOGRAPHIES**

Uwe Laufs is currently working as a researcher at Fraunhofer Institute for Industrial Engineering in Stuttgart. He has been involved in several research projects as well as projects directly funded by industrial customers. His areas of expertise and research interests include software engineering and semantic web technologies. In the area of multi-touch software engineering, he currently works on several projects that aim at the realization of multi-touch business applications.

Christopher Ruff is currently working at Fraunhofer Institute for Industrial Engineering IAO in Stuttgart, Germany. His areas of expertise and research interests include Software Engineering with a special focus on Multi-Touch Software Engineering. Christopher Ruff is lead developer of the MT4j framework. He has several years experience working with multi-touch technologies and has developed several applications ranging from multi-touch showcases to collaborative business applications.

Jan Zibuschka is currently a researcher at Fraunhofer Institute for Industrial Engineering (IAO) in Stuttgart, Germany. He has 10 years of experience in the fields of ambient and pervasive computing, ranging from mobile phones to big ambient displays. He worked on several international research projects in the area, contributing to both research prototypes and product development. His main areas of interest are viable system design, security and privacy. He started his work in the area at Fraunhofer IPSI, and is currently working at Fraunhofer IAO, after doing time at the T-Mobile Chair for m-Business and Multilateral Security.